\documentstyle[12pt,twoside,fleqn,espcrc1]{article}
\title{Quantum decoherence and an adiabatic process in macroscopic
and mesoscopic systems}

\author{Stefan V. Mashkevich\address{Institute for Theoretical
        Physics, National academy of sciences of Ukraine, \\
        252143 Kiev, Ukraine} and
        Vladimir S. Mashkevich\address{Institute of
        Physics, National academy of sciences of Ukraine, \\
        252028 Kiev, Ukraine}}

\begin{document}
\maketitle

\begin{abstract}
Quantum decoherence is of primary importance for relaxation
to an equilibrium distribution and, accordingly, for
equilibrium processes. We demonstrate how coherence breaking
implies evolution to a microcanonical distribution
(``microcanonical postulate'') and, on that ground, consider
an adiabatic process, in which there is no thermostat.
We stress its difference from a zero-polytropic process, i.e.,
a process with zero heat capacity but involving a thermostat.
We find the distribution for the adiabatic process and show that
(i) in the classical limit this distribution is canonical,
(ii) for macroscopic systems, the mean values of energy for
adiabatic and zero-polytropic processes are the same, but its
fluctuations are different, and (iii) in general, adiabatic and
zero-polytropic processes are different, which is particularly
essential for mesoscopic systems; for those latter,
an adiabatic process is in general irreversible.
\end{abstract}
\vspace*{0.5cm}

One of the most important manifestations of quantum decoherence
in the nature is in the relaxation of a thermodynamic system to an
equilibrium state.
Indeed, from the point of view of Schr\"odinger dynamics,
any eigenstate of the Hamiltonian is stationary, i.e.,
the system remains in such a state forever.
Experience shows, nevertheless, that this is not what happens
in a thermodynamic system. In such a system, each level being
highly degenerate, an arbitrary eigenvector of the Hamiltonian
can be written as $\psi_E = \sum_j c^{(j)} \psi^{(j)}_E$,
where $\{\psi^{(j)}_E\}$ is a basis in the subspace of
eigenvectors with energy $E$. What happens with time is that
coherence between $\psi^{(j)}_E$'s is broken, resulting in
a pure state $\omega_E = (\psi_E, \cdot\psi_E)$ evolving into
a mixed state $\omega_E^\prime = \sum_j w^{(j)}
(\psi^{(j)}_E, \cdot\psi^{(j)}_E)$ with $w^{(j)} = |c^{(j)}|^2$.
Repeating such a procedure again and again with different
``random'' sets of eigenvectors chosen as basis, one will end
up with
\begin{equation}
\omega^{\rm mc}_E = \sum_{j=1}^{G_E} \frac{1}{G_E}
(\psi^{(j)}_E, \cdot\psi^{(j)}_E),
\label{0}
\end{equation}
i.e., all $w^{(j)}$ being equal to $1/G_E$, where $G_E$ is the
degeneracy of the level. This is nothing but the microcanonical
distribution. Thus, decoherence implies that an arbitrary initial
eigenstate of the Hamiltonian ``smears'' over the corresponding
level, evolving thereby into a ``microcanonical state'' (\ref{0}).

(A question to be asked here is, what causes this decoherence and
how are the basis vectors chosen. A ``standard'' answer is that
decoherence occurs due to very small ``random'' interaction with
the environment, which one can never completely get rid of---so
small that it cannot change the energy of the system, but
essential in this context. The basis $\{\psi^{(j)}_E\}$, at
a given moment of time, is then the one which diagonalizes the
interaction Hamiltonian. Be that as it may,
however, a specific mechanism does not matter for the result,
which is that a thermodynamic system with definite energy
always ends up in a microcanonical state.)

A common situation is that of a system placed into a thermostat,
which is of much larger size than the system. Then the microcanonical
distribution for the complex of two implies Gibbs canonical
distribution for the system. With the temperature of the
thermostat changing, an equilibrium process will occur, in
each point of which the distribution is canonical.
However, there is one case, namely that of an adiabatic process,
i.e., a process in an adiabatically isolated system, when the
thermostat is absent. Therefore, within the framework of a
statistical theory, the adiabatic process demands a special treatment.

Indeed, in the existing statistical theory there is a process
which may be naturally called {\it zero-polytropic}, i.e.,
polytropic with zero heat capacity: the thermostat is present
(because the distribution is canonical), but $\delta Q=0$, so
that the heat capacity $c=0$, hence the name. This process is
treated as the ``true'' adiabatic process (for which $\delta Q=0$
due to the absence of the thermostat); however, not only
is the fact of that mistreatment ignored, but also the awareness
of the existence of two a priori different
processes---zero-polytropic and adiabatic---is absent.

In statistical thermodynamics, a process is characterized by a
family of distributions for the points of the curve
corresponding to the process. In each point of a zero-polytropic
curve the distribution is canonical, but for an adiabatic curve
this is by no means obvious. The principal problem is whether the
distributions for adiabatic and zero-polytropic processes
coincide. In considering the problem, the main point is to find
the law that the adiabatic distribution obeys. For this an
evolution equation is needed, which should in principle follow
from the underlying dynamical equations. The equations,
generally speaking, are not known. We use the following
approach to the problem.

As mentioned above, due to quantum decoherence,
a system with fixed energy, left by itself, eventually comes
into an equilibrium state characterized by the
microcanonical distribution. So we accept the microcanonical postulate
\cite{1}: When the energy is not fixed, the system left
by itself comes into an equilibrium state in which the
probability of its being in a pure state depends
only on the energy of the latter\footnote{Still, this
is a postulate and not a theorem, since there is its
explanation only, but no strict proof.}. On that basis,
we get the following picture for an adiabatic process.

The probability corresponding to a certain energy level is
constant so far as there is no crossing of this level with
another. On the other hand, at level crossings probabilities
equalize, due to the microcanonical postulate.
(In other words, whenever the energies of two levels
become equal, decoherence occurs between the states
belonging to those levels, ``mixing up'' the corresponding
states in the above-described manner.)
This equalization, in the general case, leads to
an entropy increase, so that an adiabatic process would not
coincide with the corresponding zero-polytropic one (for which
$dS=\delta Q/T=0$ by definition).
The total entropy increase, as can be shown, tends to zero together
with the level spacing, because the difference of probabilities
before crossing is first-order in this spacing, and the entropy
increase in one act of crossing is consequently second-order.
Therefore for a mesoscopic system, the
difference between the two processes is essential:
the lesser the system is the more the entropy increase
in an adiabatic process is tangible.
That is, an adiabatic process in a mesoscopic system
is in general irreversible.
On the other hand,
in the quasicontinuous spectrum approximation,
which is good for a macroscopic system, entropy is conserved for
the adiabatic process as well. In this approximation, the
probability distribution obeys a wave equation, the wave
velocity being expressed in terms of the density of states:
\begin{equation}
\frac{\partial w(\varepsilon,a)}{\partial a}=u(\varepsilon,
a)\frac{\partial w(\varepsilon,a)}{\partial\varepsilon},
\label{1}
\end{equation}
\begin{equation}
u(\varepsilon,a)=\frac{1}{G(\varepsilon,a)}\int\limits_0
^\varepsilon\frac{\partial G(\varepsilon',a)}{\partial a}
d\varepsilon',
\label{2}
\end{equation}
where $a$ is an external parameter,
$w(\varepsilon,a)$ is the probability of the system
being in a pure state with energy $\varepsilon$ (but not the
probability density), $G(\varepsilon,a)$ is the density of
states. The derivation of eq.~(\ref{1}) in full details is
presented in our paper \cite{2}. Given an initial
distribution $w(\varepsilon,a_0)$, one can solve
the wave equation and find the distribution at any point
of an adiabatic process.

By constructing the solution to the wave equation in the
standard way, it is seen that, the initial distribution
being canonical, the adiabatic distribution is canonical
as well (thus coinciding with the zero-polytropic
one) if $c_a$, the heat capacity at constant $a$,
is $a$ and $T$ independent. Such a system is ``canonical''
in the sense that for it the canonical distribution
is retained despite the absence of the thermostat.
This property for $c_a$ holds in the classical limit
for a system with quadratic degrees of freedom.

For a general case, an adiabatic process differs from the
relevant zero-polytropic one. Specifically,
the relative difference of the mean values of
energy for the two processes is of the order of
the inverse number of particles in the system. Therefore
in the thermodynamic limit, i.e., for a macroscopic system,
these mean values coincide. On the other hand, the
fluctuations of energy differ essentially, namely,
\begin{equation}
\Delta E=\left\{
\begin{array}{ll}
\sqrt{c_a(a,T)}T\quad & \mbox{for the zero-polytropic process},\\ \\
\sqrt{c_a(a_0,T_0)}T\quad & \mbox{for the adiabatic process},
\end{array}
\right.
\label{3}
\end{equation}
where the subscript $0$ refers to an initial state, and $T$ for
the adiabatic process means the temperature for the relevant
zero-polytropic one. This difference in fluctuations should
in principle be observable.

For a mesoscopic system, however, the difference
between the adiabatic and zero-poly\-tropic processes manifests
itself not only in the energy fluctuations, but in the mean
values of energy as well. Furthermore, the zero-polytropic
process for any system is reversible, but the adiabatic
one for a mesoscopic system is accompanied by an entropy
increase and is thereby irreversible. Thus it is mesoscopic
systems where experimental investigations on the two processes
would be most fruitful.

In conclusion, let us dwell on the estimate of the process speed
necessary for the results to take place. For the zero-polytropic
process, there is the usual condition that the process has to be
quasistatic, which means there has to be enough time for
equilibrium to be established at every moment. The same has to
hold for the adiabatic process; enough time has to be provided
for the equalization of probability to be accomplished at each
level crossing. Besides, the conditions of applicability of the
adiabatic theorem have to be fulfilled, which means that the
process has to be slow enough for the state to ``follow'' the
change of external parameters, in the quantum-mechanical sense.

\end{document}